\documentclass[aps,pra,floatfix,twocolumn]{revtex4}
\usepackage[dvips]{graphicx}

\usepackage{longtable}
\usepackage{dcolumn}
\usepackage[dvips]{graphicx}
\usepackage{bm}
\usepackage{bbm}

\usepackage{times}
\usepackage{nicefrac}
\usepackage{amsmath}
\usepackage{amsfonts}
\usepackage{amssymb}
\usepackage{amsthm}

\newcolumntype{.}{D{x}{}{-1}}

\usepackage{color}

\newcommand{\vare}{\varepsilon}

\newcommand{\bfp}{{\bm {p}}}

\newcommand{\lbr}{\langle}
\newcommand{\rbr}{\rangle}

\newcommand{\Za}{Z\alpha}

\begin{document}

\title{Nuclear recoil corrections to the Lamb shift of hydrogen and light hydrogen-like ions}

\author{V. A. Yerokhin} \affiliation{Center for Advanced Studies,
        Peter the Great St.~Petersburg Polytechnic University, Polytekhnicheskaya 29,
        St.~Petersburg 195251, Russia}
\author{V. M. Shabaev} \affiliation{Department of Physics, St. Petersburg State University,
7/9 Universitetskaya naberezhnaya, St. Petersburg 199034, Russia}

\begin{abstract}

Accurate calculations of the nuclear recoil effect to the Lamb shift of hydrogen-like atoms are
presented. Numerical results are reported for the $ns$ states with $n \leq 5$ and for the
$2p_{1/2}$ and $2p_{3/2}$ states. The calculations are performed to the first order in the
electron-nucleus mass ratio and to all orders in the nuclear binding strength parameter $\Za$
(where $Z$ is the nuclear charge number and $\alpha$ is the fine structure constant). The
obtained results provide accurate predictions for the higher-order remainder beyond the known
$\Za$-expansion terms. In the case of hydrogen, the remainder was found to be much larger than
anticipated. This result resolves the previously reported disagreement between the numerical
all-order and the analytical $\Za$-expansion approaches for the nuclear recoil effect in the
hydrogen Lamb shift.

\end{abstract}


\maketitle

\section{Introduction}

Hydrogen atom and hydrogen-like ions are the examples of the most fundamental physical systems.
Their simplicity makes them an ideal testing ground for extending the theory based on the
principles of quantum electrodynamics (QED) up to the utmost precision \cite{yerokhin:15:Hlike}.
The theory of the hydrogen Lamb shift is of particular importance because of its connections with
the determination of the Rydberg constant \cite{mohr:12:codata}, and also in view of a large
(7$\sigma$) unexplained difference of the proton charge radius as extracted from the muonic
hydrogen and the usual (electronic) hydrogen, known as the the proton charge radius puzzle
\cite{pohl:10,antognini:13}. One of the explanation of this puzzle might be a yet undiscovered
problem in the theory of the electronic hydrogen. For this reason, investigations of possible
inconsistencies in the hydrogen theory are of particular importance today.

The nonrelativistic theory of the nuclear recoil effect in a two-body system such as the
hydrogen-like atom is very simple. The nonrelativistic nuclear recoil can be accounted for to all
orders in the electron-to-nucleus mass ratio $m/M$ by introducing the reduced mass $m_r = m
M/(m+M)$ in the one-electron Schroedinger equation. The lowest-order relativistic recoil correction
can be derived from the Breit equation and is also well known \cite{bethesalpeter}. The fully
relativistic theory of the nuclear recoil, however, is highly nontrivial and can be formulated only
within the framework of quantum electrodynamics (QED).

Early studies of the QED nuclear recoil effect were performed within the approach based on the
expansion in the nuclear binding strength parameter $\Za$ \cite{salpeter:52}. The complete formulas
for the nuclear recoil effect to first order in $m/M$ and to all orders in $Z\alpha$ were first
derived by one of us \cite{shabaev:85,shabaev:88} (see also \cite{shabaev:98:rectheo}) and later
confirmed by other authors \cite{yelkhovsky:94:xxx,pachucki:95,adkins:07}. Numerical calculations
to all orders in $\Za$ were reported in
Refs.~\cite{artemyev:95:pra,artemyev:95:jpb,shabaev:98:jpb}. The results of these calculations
agreed well with the first terms of the $\Za$ expansion \cite{salpeter:52,pachucki:95,golosov:95}.
However, a  disagreement was later observed for the higher-order $\Za$ expansion terms
\cite{pachucki:99:prab,melnikov:99}. The difference between the all-order and the $\Za$-expansion
results contributed 0.7~kHz to the hydrogen $1s$ Lamb shift and was the source of the
second-largest theoretical error in the theoretical prediction \cite{mohr:12:codata}.

In the present work we report a high-precision non-perturbative (in $\Za$) calculation of the
nuclear recoil effect to the Lamb shift of energy levels in hydrogen-like atoms with $Z \le 10$.
Our first results for $n=1$ and $n=2$ states were published in Ref.~\cite{yerokhin:15:recprl}. In
the present paper we extend our calculations to a larger range of the nuclear charge numbers and to
higher excited states and describe details of the calculation.

Relativistic units ($\hbar = c = 1$) are used throughout the paper.

\section{Theory}

The recoil correction to the Lamb shift of hydrogen-like atoms, to first order in $m/M$ but to all
orders in $Z\alpha$ can be represented as a sum of four terms,
\begin{align}\label{eq:01}
\Delta E_{\rm rec} = \Delta E_{\rm L} + \Delta E_{\rm C} + \Delta E_{\rm tr(1)} + \Delta E_{\rm tr(2)}\,,
\end{align}
where $\Delta E_{\rm L}$ (the {\em low-energy part}) is the recoil correction as can be derived
from the Breit equation, $\Delta E_{\rm C}$ (the {\em Coulomb part}) is the QED recoil correction
induced by the exchange of arbitrary number of virtual Coulomb photons between the electron and the
nucleus, $\Delta E_{\rm tr(1)}$ and $\Delta E_{\rm tr(2)}$ (the {\em one-transverse-photon} and
{\em two-transverse-photons} parts, respectively) are the QED recoil corrections induced by the
exchange of one (respectively, two) transverse photon(s) and arbitrary number of virtual Coulomb
photons between the electron and the nucleus. The low-energy part $\Delta E_{\rm L}$ contains the
complete result to orders $(\Za)^2\,m/M$ and $(\Za)^4\,m/M$ and partial results for the
higher-order (in $\Za$) corrections. The remaining terms $\Delta E_{\rm C}$, $\Delta E_{\rm
tr(1)}$, and $\Delta E_{\rm tr(2)}$ induce contributions to orders $(\Za)^5\,m/M$ and higher.

In the following, we will first consider the case where the nucleus is considered to be the point
source of the Coulomb field. Additional corrections arising because of the finite nuclear charge
distribution will be addressed separately in the second part of the section.

\subsection{Point nucleus}

For the point nucleus, the low-energy part of the recoil effect $\Delta E_{\rm L}$ can be derived
from the Breit equation. It is given by \cite{shabaev:85}
\begin{align} \label{eq:02}
\Delta E_{\rm L} =& \frac{1}{2M} \langle a \bigl| \bfp^2
- \boldsymbol{D}(0) \cdot \boldsymbol{p} - \boldsymbol{p} \cdot \boldsymbol{D}(0) | a
\rangle\,,
\end{align}
where $\bfp$ is the electron momentum operator, $D_j (\omega) = -4\pi \alpha Z \alpha_i D_{ij}
(\omega, r)$, $\alpha_i$ are the Dirac matrices, and $D_{ij} (\omega, r)$ is the transverse part of
the photon propagator in the Coulomb gauge,
\begin{equation}
D_{ij}(\omega, r) = -\frac{1}{4\pi} \Bigg [ \frac{\mathrm{exp}(i|\omega
|r)}{r}\,\delta_{ij}  + \nabla_i \nabla_j \frac{\mathrm{exp}(i|\omega |r) - 1}{\omega^2 r} \Bigg ]\,.
\label{eq:propagator}
\end{equation}
Equation~(\ref{eq:02}) can be calculated analytically and cast in a very simple form
\cite{shabaev:85},
\begin{align} \label{eq:02a}
\Delta E_{\rm L} =&\ \frac{m^2 - \vare_a^2}{2M}\,,
\end{align}
where $\vare_a$ is the Dirac energy of the reference state.

The corrections $\Delta E_{\rm C}$, $\Delta E_{\rm tr,1}$, and $\Delta E_{\rm tr(2)}$ in
Eq.~(\ref{eq:01}) are derived within the QED theory \cite{shabaev:85,
shabaev:88,yelkhovsky:94:xxx,pachucki:95, shabaev:98:rectheo,adkins:07}. The result for the Coulomb
part is
\begin{align}
\Delta E_{\rm C} =
&\
\frac{2\pi i}{M}
\int_{-\infty}^\infty \mathrm{d}\omega\,
    \delta_{+}^2 (\omega) \,
    \langle a | [\boldsymbol{p}, V] \, G(\omega + \vare_a)\,  [\boldsymbol{p}, V] | a \rangle\,,
\label{eq:dE_C}
\end{align}
where $\delta_{+} (\omega) = i/(2\pi)/(\omega+i0)$, $V(r) = -\Za/r$ is the nuclear Coulomb
potential, $G(\omega) = 1/[\omega - H(1-i0)]$ is the relativistic Coulomb Green function, $H =
\boldsymbol{\alpha} \cdot \boldsymbol{p} + \beta m + V$ is the Dirac-Coulomb Hamiltonian, and $[.\
,\!\ .]$ denotes commutator. The integration over $\omega$ in Eq.~(\ref{eq:dE_C}) can be carried
out analytically, yielding
\begin{align}
\Delta E_{\rm C} = -\frac1M\, \sum_{\vare_n < 0} \lbr a |\boldsymbol{p}|n\rbr \lbr n |\boldsymbol{p}|a\rbr\,,
\label{eq:dE_C2}
\end{align}
where the summation over $n$ is extended over the negative-energy part of the Dirac spectrum and
the scalar product is implicit.

The one-transverse-photon part $\Delta E_{\rm tr(1)}$ is induced by the exchange of one transverse
and arbitrary number of Coulomb photons between the electron and the nucleus. The result is
\begin{align} \label{eq:03}
\Delta E_{\rm tr(1)} = &\
 - \frac{1}{M} \int_{-\infty}^\infty \mathrm{d}\omega\, \delta_{+} (\omega)\,
    \langle a | \bigl\{ [\boldsymbol{p}, V] \,G(\omega + \vare_a)\boldsymbol{D}(\omega)\,
 \nonumber\\ &
    - \boldsymbol{D}(\omega)\, G(\omega + \vare_a) \, [\boldsymbol{p}, V] \big\} | a \rangle\,.
\end{align}
The two-transverse-photons part $\Delta E_{\rm tr(2)}$ is induced by the exchange of two transverse
and arbitrary number of Coulomb photons between the electron and the nucleus. The result is
\begin{align}\label{eq:04}
\Delta E_{\rm tr(2)} &=& \frac{i}{2\pi M} \int_{-\infty}^\infty
\mathrm{d}\omega \, \langle a | \boldsymbol{D}(\omega) \, G(\omega + E_a) \, \boldsymbol{D}(\omega) | a \rangle\,.
\end{align}

The QED part of the recoil effect can be conveniently parameterized in terms of a slowly-varying
dimensionless functions $P(\Za)$,
\begin{align}\label{rec:1}
\Delta E_{\rm C} + \Delta E_{\rm tr(1)} + \Delta E_{\rm tr(2)} = \frac{m^2}{M}\, \frac{(\Za)^5}{\pi\, n^3}\,P(\Za)\,,
\end{align}
where $n$ is the principal quantum number of the state under consideration. For low-$Z$ atoms, the
function $P(\Za)$ can be expanded in a series over the parameter $\Za$, which is of the form
\begin{align} \label{PZa}
P(\Za) = &\ \ln (\Za)^{-2}\,D_{51} + D_{50}
  \nonumber \\ &
 + (\Za)\, D_{60}+ (\Za)^2\,G_{\rm rec}(\Za)\,,
\end{align}
where $D_{ij}$ are the coefficients and $G_{\rm rec}(\Za)$ is the higher-order remainder containing
all higher orders in $\Za$. The coefficients of the $\Za$ expansion are
\cite{salpeter:52,pachucki:95,golosov:95}
\begin{align}
D_{51}  = &\ \frac13\,\delta_{l,0}\,,\\
D_{50}  = &\ \biggl[ -\frac83\,\ln k_0(n,l) + d_{50}\biggr] \,,\\
D_{60}  = &\ \left( 4\ln 2-\frac{7}{2}\right)\pi\,\delta_{l,0}
  \nonumber \\ &
+ \left[ 3- \frac{l(l+1)}{n^2}\right]\, \frac{2\pi(1-\delta_{l,0})}{(4l^2-1)(2l+3)}\,,
\end{align}
where $\ln k_0(n,l)$ is the Bethe logarithm, whose numerical values for the states of the current
interest are \cite{drake:90}
\begin{align}
\ln k_0(1s) & = 2.984\,128\,556\,,\\
\ln k_0(2s) & = 2.811\,769\,893\,,\\
\ln k_0(3s) & = 2.767\,663\,612\,,\\
\ln k_0(4s) & = 2.749\,811\,840\,,\\
\ln k_0(5s) & = 2.740\,823\,728\,,\\
\ln k_0(2p) & =-0.030\,016\,709\,.
\end{align}
The values of the coefficients $d_{50}(n,l)$ for these states are
\begin{align}
d_{50}(ns) &\ = \frac{41}{9} + \frac{14}{3}\left[ \ln \frac{2}{n} + \sum_{i = 1}^n \frac1i - \frac1{2n}\right]\,,\\
d_{50}(2p) &\ = -\frac{7}{18}\,.
\end{align}

The $\Za$ expansion of the higher-order remainder reads
\begin{align} \label{Zaho}
G_{\rm rec}(\Za) &\, = \ln^2 (\Za)^{-2}\,D_{72} + \ln (\Za)^{-2}\,D_{71} + D_{70} + \ldots\,,
\end{align}
where only the double logarithmic contribution is presently known
\cite{pachucki:99:prab,melnikov:99}
\begin{align} \label{D72}
D_{72}  = &\  -\frac{11}{60}\,\delta_{l,0}\,.
\end{align}

\subsection{Extended nucleus}

The low-energy part of the recoil correction for the case of an extended nuclear charge was derived
in Refs.~\cite{grotch:69,borie:82} (see also \cite{aleksandrov:15}),
\begin{align} \label{eq:elexact}
\Delta E_{\rm L} = &\ \frac{1}{2M}\, \lbr a | \bigl[\vare_{a}^2 - m^2
\nonumber \\ &
    - 2m \beta V(r) - W^\prime (r) V^\prime(r) - V^2(r) \bigr] | a \rbr\,,
\end{align}
where \begin{eqnarray} \label{eq:V}
V(r) &=& -\Za\int \mathrm{d}\boldsymbol{r}^\prime \frac{\rho(\boldsymbol{r}^\prime)}{|\boldsymbol{r} - \boldsymbol{r}^\prime|}, \label{eq:V}\\
W(r) &=& -\Za\int \mathrm{d}\boldsymbol{r}^\prime \rho(\boldsymbol{r}^\prime)|\boldsymbol{r} -
\boldsymbol{r}^\prime|, \label{eq:W}
\end{eqnarray}
$V^{\prime}(r) = {\rm d}V(r)/{\rm d}r$, $W^{\prime}(r) = {\rm d}W(r)/{\rm d}r$,  and $\rho
(\boldsymbol{r})$ is the density of the nuclear charge distribution $\big ( \int
\mathrm{d}\boldsymbol{r} \rho (\boldsymbol{r}) = 1\big )$.

The Coulomb part of the QED recoil correction $\Delta E_{\rm C}$ for an extended nuclear charge is
given by the same formula (\ref{eq:dE_C}) as for the point-nucleus case, with $V(r)$ being the
extended-nucleus potential (\ref{eq:V}). Exact expressions for the one-transverse-photon part
$\Delta E_{\rm tr(1)}$ and the two-transverse-photons part $\Delta E_{\rm tr(2)}$ for the extended
nucleus case are not yet known. In the present work, we will use the expressions (\ref{eq:03}) and
(\ref{eq:04}) derived for the point nucleus but evaluate these expressions with the
extended-nucleus wave functions, energies, electron propagators, and nuclear potential. The same
treatment was presented earlier in Refs.~\cite{shabaev:98:recground,shabaev:98:ps}.  The
uncertainty introduced by this approximation will be discussed in Sec.~\ref{sec:results}.

We are interested in the recoil correction induced by the finite nuclear size (fns), so we take a
difference between the results obtained with an extended nucleus and with the point nucleus,
\begin{align}
E_{\rm fns, rec} = E_{\rm rec}({\rm ext}) - E_{\rm rec}({\rm pnt})\,,
\end{align}
where ${\rm ext}$ and ${\rm pnt}$ refer to the extended and the point nuclear distributions,
respectively.

For the low-$Z$ atoms, it is customary \cite{mohr:12:codata,yerokhin:15:Hlike} to account for a
part of the recoil fns effects by introducing the reduced mass prefactor $(M/(m+M))^3$ in the
expression for the fns correction. To the first order in $m/M$, such correction is given by
\begin{align} \label{eq:101}
E_{\rm fns, rm} = -3\,\frac{m}{M}[\vare_a({\rm ext})-\vare_a({\rm pnt})]\,,
\end{align}
where $\vare_a({\rm ext})$ and $\vare_a({\rm pnt})$ are the eigenvalues of the Dirac equation with
the extended and the point nuclear potentials, respectively.

In the present work, we will identify the higher-order fns recoil correction that is {\em beyond}
the reduced-mass part (\ref{eq:101}) and parameterize it in terms of the function $\delta_{\rm
fns}P$,
\begin{align}
E_{\rm fns, rec} = E_{\rm fns, rm} + \frac{m^2}{M}\, \frac{(\Za)^5}{\pi\, n^3}\,\delta_{\rm fns}P\,.
\end{align}
We would like to draw the reader's attention to the fact that $\delta_{\rm fns}P$ includes the fns
contribution from $\Delta E_{\rm L}$.

\section{Numerical calculation}

The general scheme of the calculation was described previously in Ref.~\cite{artemyev:95:pra}. An
important issue in the computation of Eqs.~(\ref{eq:dE_C})-(\ref{eq:04}) is the adequate numerical
representation of the Dirac-Coulomb Green function $G(\omega)$. The Dirac-Coulomb Green function is
known in a form of the partial-wave expansion over the angular momentum-parity quantum number
$\kappa$ (see, e.g., Ref.~\cite{mohr:98} for details). After all angular-momentum selection rules
are taken into account, only a few of the partial-wave contributions of the Green function survive
(two for the $j = 1/2$ reference states and three for the $j = 3/2$ reference states). The
resulting expressions were evaluated by summing over the spectrum of the radial Dirac equation with
the help of the finite basis set constructed with $B$-splines. For the point nuclear model we used
the standard variant of the $B$-spline method \cite{johnson:88}. The calculations for the extended
nucleus were performed by the Dual kinetic balance method \cite{shabaev:04:DKB}.

The main technical problem of the previous calculations \cite{artemyev:95:pra,shabaev:98:jpb} that
limited the  numerical accuracy in the low-$Z$ region was lack of convergence with increase of the
size of the basis set $N$. In the present investigation, we found out that this effect was caused
by numerical instabilities associated with limitations of the standard double-precision
(approximately 16 digits) arithmetics. In the present work we implemented the procedure of solving
the Dirac equation with the $B$-splines basis set in the quadruple-precision (approximately
32-digit) arithmetics. After that we were able to achieve a clear convergence pattern of the
calculated results when the size of the basis set was increased. The largest basis size used in
actual calculations was $N = 250$. The numerical uncertainty of the obtained results was estimated
by changing the size of the basis set by 30-50\% and by increasing the number of integration points
in numerical quadratures.

Calculations for the extended nucleus were performed with two models of the nuclear charge
distribution, the Gauss model and the homogeneously charged sphere model. We did not use the Fermi
model, which is commonly used in calculations of heavy and medium-$Z$ atoms, since this model is
not suitable for very light nuclei. The Gauss distribution of the nuclear charge reads
\begin{align} \label{1a}
\rho_{\rm Gaus}(r) = \left( \frac{3}{2\pi R^2 }\right)^{3/2}\,\exp\left(-\frac{3\, r^2}{2R^2}\right)\,,
\end{align}
where $R$ is the root-mean-square radius of the nuclear charge distribution. The homogeneously
charged sphere distribution is given by
\begin{align} \label{1}
\rho_{\rm Sph}(r) = \frac{3}{4\pi R^3_{\rm Sph}}\,\theta(R_{\rm Sph}-r)\,,
\end{align}
where $\theta(r)$ is the Heaviside step function and $R_{\rm Sph} = \sqrt{5/3}\,R$. We estimate the
nuclear model dependence of our calculations by comparing the results obtained for these two
nuclear models.

In our calculations, we had to numerically evaluate integrals of the form
\begin{align} \label{eq:001}
F(\Delta) = \int_0^{\infty}dy\, \frac{\Delta}{\Delta^2+y^2}\,f(y)\,,
\end{align}
where $\Delta = \vare_a-\vare_n$ is the energy difference of the reference state and the virtual
state and $f(y)$ is a smooth function of $y$. The integration over $y$ was performed numerically by
splitting the interval $(0,\infty)$ into subintervals, making suitable change of variables and
applying the Gauss-Legendre quadratures. Special care had to be taken in performing numerical
integrations when $|\Delta|$ happens to be small, since the integrand has a rapidly changing
structure at $y \sim |\Delta|$. In such cases, we represent $F(y)$ as a sum of 3 terms,
\begin{align} \label{eq:002}
F(\Delta) = &\ f(0)\,\frac{\Delta}{|\Delta|}\,\arctan\frac{\Za}{|\Delta|}
  \nonumber \\ & +
\int_0^{\Za}dy\, \frac{\Delta}{\Delta^2+y^2}\,[f(y)-f(0)]
  \nonumber \\ & +
\int_{\Za}^{\infty}dy\, \frac{\Delta}{\Delta^2+y^2}\,f(y)\,.
\end{align}
Taking into account that for small $x$, $\arctan(1/x) = \pi/2-x+\ldots$, it is easy to see that
Eq.~(\ref{eq:002}) has a smooth and numerically safe transition to the limit $\Delta\to 0$, in
contrast to the original expression (\ref{eq:001}). In order to calculate the second term in the
right-hand side of Eq.~(\ref{eq:002}), we first store the (slowly varying) function $f(y)-f(0)$ on
a grid and then compute the integral over $y$ numerically with obtaining function $f(y)-f(0)$ by a
polynomial interpolation. The third term in the right-hand side of Eq.~(\ref{eq:002}) does not
represent any problems and is evaluated in the standard way.

\section{Results and discussion}
\label{sec:results}

For the point nuclear model, our numerical results for the $n= 1$ and $n= 2$ states are presented
in Table~\ref{tab:pnt} and those for the $ns$ states with $n = 3 \ldots 5$ in
Table~\ref{tab:pnt:2}. Table~\ref{tab:pnt} presents also a comparison with the previous numerical
 and $\Za$-expansion calculations. Generally, we
find very good agreement with previous numerical results
\cite{artemyev:95:pra,artemyev:95:jpb,shabaev:98:jpb}. The only exception is the $2s$ state and $Z
= 1$, for which a small deviation is found that was caused by a minor mistake in the previous
calculation. At the same time, we observe a strong contrast between the all-order results for the
higher-order remainders $G_{\rm rec}(1s)$ and $G_{\rm rec}(2s)$ and the corresponding
$\Za$-expansion values \cite{pachucki:99:prab,melnikov:99}. We recall that the $\Za$-expansion
results for $G_{\rm rec}$ include only the double-log contribution $D_{72} \ln^2 (\Za)^{-2}$ and
neglect the higher-order terms. For hydrogen, $\ln (1\alpha)^{-2} \approx 10$ is a large parameter.
For this reason, the leading logarithmic approximation is often used for estimating the tail of the
$\Za$ expansion, with a typical estimate of uncertainty of 50\% \cite{mohr:05:rmp}.

\begin{figure*}
\centerline{
\resizebox{0.85\textwidth}{!}{%
  \includegraphics{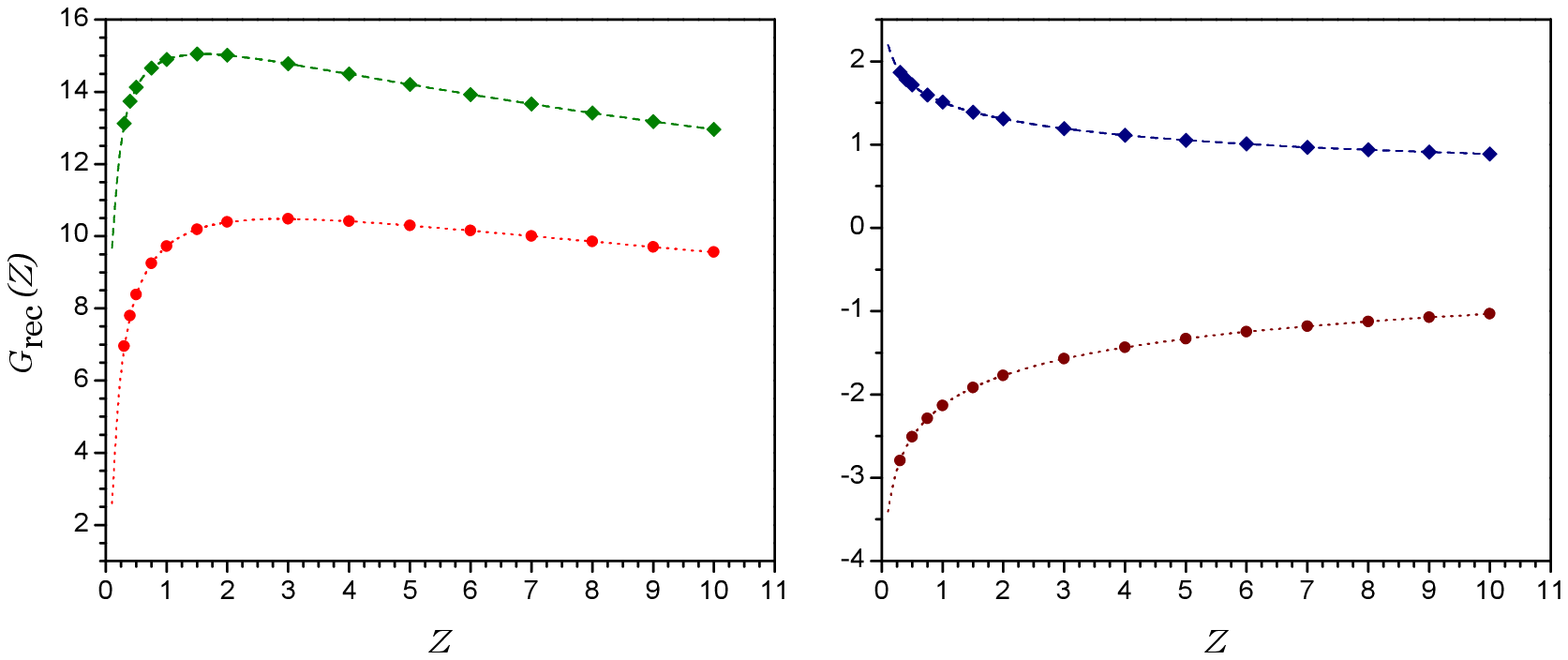}
}}
 \caption{(Color online)
Higher-order recoil correction $G_{\rm rec}(Z)$ for the $1s$ state (left graph, dots, red), the $2s$ state (left graph, diamonds, green),
the $2p_{1/2}$ state (right graph, diamonds, blue), the $2p_{3/2}$ state (right graph, dots, brown).
 \label{fig:Grec}}
\end{figure*}

In order to perform a detailed analysis of the seeming discrepancy with the $\Za$ expansion
results, we performed our calculations for a series of nuclear charges including fractional $Z$
values as low as $Z = 0.5$. The results obtained for the higher-order remainder $G_{\rm rec}(\Za)$
are plotted in Fig.~\ref{fig:Grec}. We discover a rapidly changing structure at very low values of
$Z$. Most remarkably, the bending of the curve is practically undetectable for $Z \ge 2$. In order
to access such a structure in an all-order calculation, one needs to achieve a very high numerical
accuracy at very low (and fractional) values of $Z$.

\begin{table*}
\caption{Nuclear recoil correction for the $n = 1$ and $n = 2$ states and the point nuclear model, expressed in terms of
$P(\Za)$ and $G_{\rm rec}(\Za)$, $1/\alpha = 137.0359895$.
\label{tab:pnt}
}
\begin{ruledtabular}
  \begin{tabular}{l........}
& \multicolumn{2}{c}{$1s$}
& \multicolumn{2}{c}{$2s$}
& \multicolumn{2}{c}{$2p_{1/2}$}
& \multicolumn{2}{c}{$2p_{3/2}$}
 \\
& \multicolumn{1}{c}{$P(\Za)$}
& \multicolumn{1}{c}{$G_{\rm rec}(\Za)$}
& \multicolumn{1}{c}{$P(\Za)$}
& \multicolumn{1}{c}{$G_{\rm rec}(\Za)$}
& \multicolumn{1}{c}{$P(\Za)$}
& \multicolumn{1}{c}{$G_{\rm rec}(\Za)$}
& \multicolumn{1}{c}{$P(\Za)$}
& \multicolumn{1}{c}{$G_{\rm rec}(\Za)$}
 \\
 \hline\\[-3pt]
  0.50 & 5.899\,x9336\,(2) & 8.38x\,(2) &   6.624\,9x463\,(2) & 14.13x\,(1)  & -0.305\,0x00\,61\,(1) & 1.715\,x4\,(8)  & -0.305\,x056\,85\,(1) & -2.509x1\,(8) \\[2pt]
  0.75 & 5.625\,x6199\,(2) & 9.24x6\,(6) &  6.350\,7x183\,(2) & 14.66x0\,(5) & -0.303\,0x65\,24\,(1) & 1.594\,x6\,(3)  & -0.303\,x181\,53\,(1) & -2.287x7\,(3)\\[2pt]
  1    & 5.429\,x9035\,(2) & 9.72x0\,(3) &  6.155\,1x155\,(2) & 14.89x9\,(3) & -0.301\,1x22\,17\,(1) & 1.509\,x7\,(2)  & -0.301\,x316\,16\,(1) & -2.133x3\,(2)  \\
       & 5.429\,x90\,(3)^a     &         &  6.154\,8x3\,(5)^a     &          & -0.301\,1x2^a \\
       & 5.430\,x(2)^b          &        &  6.155\ (x1)^b          &         & -0.301\,1x^b                &         & -0.301\,x3^b\\
       & 5.428\,x441\,^c    & -17.75x\,^c & 6.153\,3x77\,^c &-17.75x\,^c   &  -0.301\,2x03\,^c      &                 & -0.301\,x203\,^c  \\[2pt]
  1.50 & 5.151\,x9588\,(2) & 10.19x2\,(1) & 5.877\,4x764\,(2) & 15.04x3\,(1) & -0.297\,2x15\,04\,(1) & 1.390\,x77\,(8) & -0.297\,x611\,69\,(1) & -1.919x73\,(8) \\[2pt]
  2    & 4.952\,x8246\,(3) & 10.39x0\,(1) & 5.678\,7x451\,(3) & 15.01x0\,(1) & -0.293\,2x82\,31\,(1) & 1.307\,x39\,(5) & -0.293\,x938\,24\,(1) & -1.772x02\,(5)   \\[2pt]
  3    & 4.668\,x6482\,(5) & 10.48x03\,(9) & 5.395\,6x454\,(5) & 14.78x06\,(9)& -0.285\,3x47\,72\,(1) & 1.192\,x04\,(2) & -0.286\,x671\,66\,(1) & -1.570x41\,(2)\\[2pt]
  4    & 4.464\,x0355\,(5) & 10.41x55\,(6) & 5.192\,4x455\,(5) & 14.49x26\,(6)& -0.277\,3x29\,22\,(2) & 1.112\,x68\,(2) & -0.279\,x498\,03\,(1) & -1.432x80\,(1)\\[2pt]
  5    & 4.303\,x4275\,(5) & 10.29x44\,(4) & 5.033\,5x649\,(5) & 14.20x13\,(4)& -0.269\,2x33\,36\,(2) & 1.053\,x21\,(2) & -0.272\,x405\,66\,(2) & -1.329x68\,(2)\\[2pt]
  6    & 4.170\,x9596\,(5) & 10.15x14\,(2) & 4.903\,1x236\,(5) & 13.92x16\,(2)& -0.261\,0x64\,35\,(3) & 1.006\,x42\,(2) & -0.265\,x386\,15\,(3) & -1.247x98\,(2)\\[2pt]
  7    & 4.058\,x1513\,(5) & 10.00x10\,(2) & 4.792\,6x291\,(5) & 13.65x77\,(2)& -0.252\,8x24\,98\,(4) & 0.968\,x44\,(2) & -0.258\,x433\,08\,(3) & -1.180x82\,(1)\\[2pt]
  8    & 3.959\,x9275\,(5) & 9.84x98\,(2) & 4.696\,9x974\,(5) & 13.41x00\,(2) & -0.244\,5x16\,94\,(4) & 0.936\,x96\,(1) & -0.251\,x541\,38\,(4) & -1.124x15\,(1)\\[2pt]
  9    & 3.873\,x0050\,(6) & 9.70x12\,(1) & 4.612\,9x380\,(6) & 13.17x80\,(1) & -0.236\,1x41\,15\,(6) & 0.910\,x48\,(1) & -0.244\,x706\,89\,(5) & -1.075x38\,(1)\\[2pt]
 10    & 3.795\,x1349\,(6) & 9.55x67\,(1) & 4.538\,1x963\,(6) & 12.96x04\,(1) & -0.227\,6x97\,95\,(7) & 0.887\,x99\,(1) & -0.237\,x926\,12\,(6) & -1.032x74\,(1)\\
       & 3.795\,x0\,(1)^b  &              & 4.538\,3x\,(1)^b  &               & -0.227\,7x\,^b      &                 & -0.237\,x9\,^c      \\
\end{tabular}
\end{ruledtabular}
$^a\,$ Shabaev et al. 1998 \cite{shabaev:98:jpb}\,\\
$^b\,$ Artemyev et al. 1995 \cite{artemyev:95:pra,artemyev:95:jpb}\,\\
$^c\,$ $\Za$ expansion
\end{table*}

\begin{table*}
\caption{Nuclear recoil correction for the $3s$, $4s$, and $5s$ states and the point nuclear model, expressed in terms of
$P(\Za)$ and $G_{\rm rec}(\Za)$, $1/\alpha = 137.0359895$.
\label{tab:pnt:2}
}
\begin{ruledtabular}
  \begin{tabular}{l......}
& \multicolumn{2}{c}{$3s$}
& \multicolumn{2}{c}{$4s$}
& \multicolumn{2}{c}{$5s$}
 \\
& \multicolumn{1}{c}{$P(\Za)$}
& \multicolumn{1}{c}{$G_{\rm rec}(\Za)$}
& \multicolumn{1}{c}{$P(\Za)$}
& \multicolumn{1}{c}{$G_{\rm rec}(\Za)$}
& \multicolumn{1}{c}{$P(\Za)$}
& \multicolumn{1}{c}{$G_{\rm rec}(\Za)$}
 \\
 \hline\\[-3pt]
  1    & 6.325\,x0244\,(2) & 15.2x42\,(3) & 6.391\,x2172\,(2)  & 15.1x15\,(3)  & 6.423\,x8396\,(2) & 14.9x41\,(3) \\
  2    & 5.848\,x6979\,(3) & 15.3x02\,(1) & 5.914\,x8703\,(4)  & 15.1x74\,(2)  & 5.947\,x4666\,(4) & 15.0x08\,(2) \\
  3    & 5.565\,x6623\,(5) & 15.0x44\,(1) & 5.631\,x8011\,(5)  & 14.9x17\,(1)  & 5.664\,x3557\,(5) & 14.7x56\,(1)    \\
  4    & 5.362\,x5443\,(5) & 14.7x369\,(6) & 5.428\,x6365\,(5) & 14.6x110\,(6) & 5.461\,x1345\,(5) & 14.4x540\,(6)    \\
  5    & 5.203\,x7622\,(5) & 14.4x316\,(4) & 5.269\,x7953\,(5) & 14.3x066\,(4) & 5.302\,x2222\,(5) & 14.1x527\,(4) \\
  6    & 5.073\,x4347\,(5) & 14.1x410\,(3) & 5.139\,x3965\,(5) & 14.0x170\,(3) & 5.171\,x7384\,(5) & 13.8x658\,(3)    \\
  7    & 4.963\,x0690\,(5) & 13.8x682\,(2) & 5.028\,x9475\,(5) & 13.7x451\,(2) & 5.061\,x1907\,(5) & 13.5x962\,(2)       \\
  8    & 4.867\,x5804\,(5) & 13.6x132\,(2) & 4.933\,x3637\,(5) & 13.4x911\,(2) & 4.965\,x4949\,(5) & 13.3x442\,(1) \\
  9    & 4.783\,x6782\,(6) & 13.3x749\,(1) & 4.849\,x3547\,(6) & 13.2x537\,(1) & 4.881\,x3607\,(6) & 13.1x086\,(1)    \\
 10    & 4.709\,x1073\,(6) & 13.1x520\,(1) & 4.774\,x6656\,(6) & 13.0x316\,(1) & 4.806\,x5334\,(6) & 12.8x882\,(1)
\end{tabular}
\end{ruledtabular}
\end{table*}

We now analyse our numerical results obtained for the higher-order remainder $G_{\rm rec}(\Za)$ by
fitting them to the following anzatz that incorporates the known form of the $\Za$ expansion to
order $(\Za)^7$,
\begin{align}
G_{\rm rec}(\Za) = d_{7,2}\,\ln^2(\Za)^{-2} +  d_{7,1}\,\ln(\Za)^{-2}
 + \sum_{i = 0}^n d_{7+i,0} (\Za)^i\,,
\end{align}
where $n = 2\,\ldots 4$ and $d_{i,k}$ are fitting coefficients. We use the fitted values of
$d_{7,j}$ coefficients as approximations for the $D_{7j}$ coefficients in the $\Za$ expansion
(\ref{Zaho}). The uncertainties are determined by changing of the length of the anzatz (i.e., $n$),
by varying the number of fitted data points, and also by changing between using the analytical
value of $D_{72}$ and fitting it as a free parameter. For the squared logarithmic contributions we
find the fitting results $D_{72}(1s)= -0.183\,(1)$ and $D_{72}(2s)= -0.183\,(1)$ which perfectly
agree with the analytical value $-11/60 = -0.18333\ldots$. The next two coefficients are:
\begin{align}
&\ D_{71}(1s)  = 2.919\,(10)\,,\ \ \
D_{70}(1s)  = -1.32\,(10)\,,\\
&\ D_{71}(2s)  = 3.335\,(10)\,,\ \ \
D_{70}(2s)  = -0.26\,(6)\,,\\
&\  D_{71}(2p_{1/2}) = 0.149\,(5)\,,\ \ \
D_{70}(2p_{1/2})  = -0.035\,(15)\,,\\
&\ D_{71}(2p_{3/2}) =  -0.283\,(5)\,,\ \ \
D_{70}(2p_{3/2})  = 0.685\,(20)\,.
\end{align}
The results obtained for the $D_{7i}$ coefficients rely on the equivalence of the nonperturbative
and the $\Za$ expansion approaches. This equivalence follows from the systematic derivation of the
$\Za$ expansion from the full QED within the formalism of nonrelativistic quantum electrodynamics
(NRQED) \cite{caswel_lepage:86} and was also confirmed by explicit calculations in different
physical contexts \cite{mohr:12:codata}.

We thus conclude that our all-order results are consistent with all known coefficients of the $\Za$
expansion. The deviation observed for the higher-order remainder of the $s$ states in
Table~\ref{tab:pnt} comes from the higher-order terms, whose contribution turns out to be
unexpectedly large. Specifically, the single-log coefficient $D_{71}$ is found to be 16 times
larger than the double-log coefficient $D_{72}$. As a result, the inclusion of the single-log
contribution changes drastically the $\Za$ expansion result for the higher-order recoil effect.

We now turn to the correction to the nuclear recoil effect induced by the finite nuclear size. The
numerical results for the higher-order fns recoil effect are presented in Table~\ref{tab:fns}. The
values of the rms radii of the nuclear charge distribution $R$ used in the calculations
\cite{angeli:13} are listed in the second column of the table. For $Z=1$ we performed calculations
for two values of $R$, one corresponding to hydrogen and another, to deuterium. Numerical results
obtained for two different models of the nuclear charge distribution are listed for each $Z$ in the
upper and lower lines, thus giving an opportunity to estimate the model dependence of the obtained
results.

The errors specified in Table~\ref{tab:fns} are estimations of the uncertainty of the approximation
made in calculations of $\Delta E_{\rm tr(1)}$ and $\Delta E_{\rm tr(2)}$ for an extended nucleus.
The numerical errors are much smaller and not indicated. The errors due to uncertainties of the
nuclear radii are also not listed. They might be easily accounted for separately, e.g., by a simple
estimate $(2\,\delta R/R)\,\delta_{\rm fns} P \,,$ where $\delta R$ is the uncertainty of the
radius $R$.

In order to estimate the uncertainty of the approximation, we compare the low-order part as
evaluated in two ways: first, by the exact formula (\ref{eq:elexact}), $\Delta E_{\rm L}$, and,
second, using the operators derived for a point nucleus (see Eq.~(4) of
Ref.~\cite{shabaev:98:recground}), $\Delta E_{\rm L}^{\rm appr}$. We then estimate the
approximation error as the absolute value of
\begin{align} \label{eq:201}
2\,\frac{\Delta E_{\rm L}-\Delta E_{\rm L}^{\rm appr}}{E_{\rm fns, rec} - E_{\rm fns, rm}}\left[
\Delta E_{\rm tr(1), fns} + \Delta E_{\rm tr(2), fns} \right]\,,
\end{align}
where  $\Delta E_{\rm tr(1), fns}$ and $\Delta E_{\rm tr(2), fns}$ are the fns corrections to the
one-transverse-photon and the two-transverse-photons parts, respectively. In Eq.~(\ref{eq:201}),
the numerator $\Delta E_{\rm L}-\Delta E_{\rm L}^{\rm appr}$ is the error of the approximation for
the low-order part, the denominator $E_{\rm fns, rec} - E_{\rm fns, rm}$ is the total value of the
recoil fns correction, whereas 2 is a conservative factor. We note that we cannot use $\Delta
E_{\rm L}$ in the denominator of Eq.~(\ref{eq:201}) because of large cancellations of spurious
terms between $\Delta E_{\rm L}$ and $\Delta E_{\rm C}$ \cite{shabaev:98:recground}. It might be
also mentioned that the full expressions for the two-transverse-photons fns correction should
contain contributions induced by virtual nuclear excitations \cite{shabaev:98:rectheo,salpeter:52}.
These terms need to be considered together with the nuclear polarization effect
\cite{khriplovitch:00,pachucki:07:heliumnp} and are beyond the scope of the present investigation.

\begin{table*}
\caption{Finite nuclear size recoil correction, expressed in terms of
$\delta_{\rm fns} P$. For each $Z$, the upper line corresponds to the Gauss nuclear model, whereas the second
line corresponds to the homogeneously charged sphere nuclear model. The specified uncertainty is the estimated
error of the approximation.
\label{tab:fns}
}
\begin{ruledtabular}
  \begin{tabular}{lc....}
$Z$ & $R$ [fm]
& \multicolumn{1}{c}{$1s$}
& \multicolumn{1}{c}{$2s$}
& \multicolumn{1}{c}{$2p_{1/2}$}
& \multicolumn{1}{c}{$2p_{3/2}$}
 \\
 \hline\\[-4pt]
  1    & 0.8775 & -0.000\,x1840\,(8) & -0.000\,x1840\,(8) & -0.000\,x000\,01\,(1)      & -0.000\,x000\,01\,(1) \\
       &        & -0.000\,x1851\,(8) & -0.000\,x1852\,(8) & -0.000\,x000\,01\,(1)      & -0.000\,x000\,01\,(1) \\
       &        &  0.000\,x0\,(2)^a \\[2pt]
       & 2.1424 & -0.000\,x7861\,(8) & -0.000\,x7866\,(8) & -0.000\,x000\,03\,(5) & -0.000\,x000\,04\,(4) \\
       &        & -0.000\,x7918\,(9) & -0.000\,x7923\,(9) & -0.000\,x000\,03\,(5) & -0.000\,x000\,04\,(4) \\[2pt]
  2    & 1.6755 & -0.000\,x628\,(4)  & -0.000\,x629\,(4)  & -0.000\,x000\,06\,(6) & -0.000\,x000\,04\,(4) \\
       &        & -0.000\,x632\,(4)  & -0.000\,x633\,(4)  & -0.000\,x000\,06\,(6) & -0.000\,x000\,04\,(4) \\
       &        & -0.000\,x6\,(2)^a \\[2pt]
  3    & 2.4440 & -0.001\,x282\,(12)  & -0.001\,x285\,(12)  & -0.000\,x000\,2\,(4)  & -0.000\,x000\,1\,(1) \\
       &        & -0.001\,x292\,(12)  & -0.001\,x294\,(12)  & -0.000\,x000\,2\,(4)  & -0.000\,x000\,1\,(1) \\[2pt]
  4    & 2.5190 & -0.001\,x50\,(2)   & -0.001\,x50\,(2)   & -0.000\,x000\,3\,(7)  & -0.000\,x000\,2\,(2) \\
       &        & -0.001\,x51\,(2)   & -0.001\,x51\,(2)   & -0.000\,x000\,3\,(7)  & -0.000\,x000\,2\,(2) \\[2pt]
  5    & 2.4060 & -0.001\,x56\,(2)   & -0.001\,x56\,(2)   & -0.000\,x000\,4\,(10)  & -0.000\,x000\,2\,(2) \\
       &        & -0.001\,x57\,(2)   & -0.001\,x57\,(2)   & -0.000\,x000\,4\,(10)  & -0.000\,x000\,2\,(2) \\
       &        & -0.001\,x5\,(2)^a \\[2pt]
  6    & 2.4702 & -0.001\,x77\,(4)   & -0.001\,x78\,(4)   & -0.000\,x001\,(1)  & -0.000\,x000\,2\,(4) \\
       &        & -0.001\,x78\,(4)   & -0.001\,x79\,(4)   & -0.000\,x001\,(1)  & -0.000\,x000\,2\,(4) \\[2pt]
  7    & 2.5582 & -0.002\,x02\,(4)   & -0.002\,x03\,(4)   & -0.000\,x001\,(1)  & -0.000\,x000\,2\,(6) \\
       &        & -0.002\,x04\,(4)   & -0.002\,x05\,(4)   & -0.000\,x001\,(1)  & -0.000\,x000\,2\,(6) \\[2pt]
  8    & 2.6991 & -0.002\,x35\,(6)   & -0.002\,x37\,(6)   & -0.000\,x001\,(2) & -0.000\,x000\,3\,(8) \\
       &        & -0.002\,x37\,(6)   & -0.002\,x38\,(6)   & -0.000\,x001\,(2) & -0.000\,x000\,3\,(8) \\[2pt]
  9    & 2.8976 & -0.002\,x78\,(8)   & -0.002\,x80\,(8)   & -0.000\,x002\,(3) & -0.000\,x000\,3\,(12)  \\
       &        & -0.002\,x80\,(8)   & -0.002\,x82\,(8)   & -0.000\,x002\,(3) & -0.000\,x000\,3\,(12)  \\[2pt]
  10   & 3.0055 & -0.003\,x12\,(8)   & -0.003\,x14\,(8)   & -0.000\,x003\,(3) & -0.000\,x000\,4\,(16)  \\
       &        & -0.003\,x14\,(8)   & -0.003\,x17\,(8)   & -0.000\,x003\,(3) & -0.000\,x000\,4\,(16)  \\
       &        & -0.003\,x(2)^a \\
\end{tabular}
\end{ruledtabular}
$^a\,$ Shabaev et al. 1998 \cite{shabaev:98:recground}\,\\
\end{table*}

\section{Summary}

In the present investigation we calculated the nuclear recoil correction to the Lamb shift of light
hydrogen-like atoms. The calculation is performed to the first order in the electron-nucleus mass
ratio $m/M$ and to all orders in the nuclear binding strength parameter $\Za$, both for the point
and the extended nuclear models. The results were found to be in excellent agreement with those
obtained previously within the $\Za$ expansion approach. The higher-order recoil contribution
beyond the previously known $\Za$-expansion terms was identified.

Our calculation resolves the previously reported disagreement between the numerical all-order and
the analytical $\Za$-expansion approaches and eliminates the second-largest theoretical uncertainty
in the hydrogen Lamb shift of the $1S$ and $2S$ states. The calculated values of the higher-order
recoil correction beyond the previously known $\Za$-expansion terms for hydrogen are $0.65$~kHz for
the $1S$ state and $0.08$~kHz for the $2S$ state, for the point nuclear model. The finite nuclear
size effect beyond the reduced mass shifts the above values by $-0.08$ and $-0.01$~kHz,
respectively. These results may be compared with the experimental uncertainty of $0.01$~kHz for the
$1S$-$2S$ transition \cite{parthey:11}.

The higher-order recoil corrections calculated in the present work influence the interpretation of
experimental results for the hydrogen-deuterium isotope shift \cite{parthey:10}. Specifically, our
results increase the theoretical value of the hydrogen-deuterium $1S$-$2S$ isotope shift as
reported in Ref.~\cite{jentschura:11} by $0.36$~kHz (including $0.28$~kHz from the point nucleus
and $0.08$~kHz from the finite nuclear size). These results may be compared to the experimental
uncertainty of $0.015$~kHz \cite{parthey:10} and the total theoretical uncertainty of $0.6$~kHz.
The change of the theoretical value increases the deuteron-hydrogen mean-square charge-radii
difference as obtained in Ref.~\cite{jentschura:11} by 0.00026~fm$^2$.

The results obtained in the present work demonstrate the importance of the non-perturbative (in
$\Za$) calculations as an alternative to the traditional $\Za$-expansion approach. Despite the
smallness of the parameter $\Za$ for hydrogen, $1\alpha \approx 0.0073$, the convergence of the
(semi-analytical) $\Za$ expansion is complicated by the presence of powers of logarithms. Moreover,
the predictive power of the $\Za$ expansion calculations is limited by the difficulty to reliably
estimate contributions of the uncalculated tail of the expansion. The non-perturbative
calculations, while clearly preferable over the perturbative ones, are often hampered by technical
difficulties associated with large numerical cancellations occurring in the low-$Z$ region. In
particular, technical difficulties prevented so far a direct numerical calculation of the two-loop
electron self-energy for hydrogen \cite{yerokhin:09:sese}, which is badly needed as this effect
presently determines the theoretical uncertainty in the hydrogen $1S$ and $2S$ Lamb shifts
\cite{mohr:12:codata}.

\section*{Acknowledgement}

V.A.Y. acknowledges support by the Russian Federation program for organizing and carrying out
scientific investigations and by RFBR (grant No. 16-02-00538). The work of V.M.S. was supported by
RFBR (grant No. 16-02-00334) and by SPbSU (grants No. 11.38.269.2014 and 11.38.237.2015).


\end{document}